\documentclass{aa}

\usepackage[varg]{txfonts}
\usepackage{color}

\begin{document}

\title{Decay of COSAC and Ptolemy Mass Spectra at Comet 67P/Churyumov-Gerasimenko}


\author{Harald~Kr\"uger\inst{1}\thanks{Contact address: krueger@mps.mpg.de} \and
Fred Goesmann \inst{1} \and 
Chaitanya Giri \inst{2,1}  \and 
Ian Wright \inst{3}  \and 
Andrew Morse \inst{3}   \and 
Jan Hendrik Bredeh\"oft \inst{4}  \and 
Stephan Ulamec \inst{5}    \and 
Barbara Cozzoni \inst{5}     \and
Pascale Ehrenfreund \inst{6,7}   \and
Thomas Gautier \inst{8,9}    \and   
Susan McKenna-Lawlor \inst{10}    \and
Francois Raulin \inst{11}      \and
Harald Steininger \inst{1}  \and
Cyril Szopa \inst{12}   
}    

\institute{Max-Planck-Institut f\"ur Sonnensystemforschung, Justus-von-Liebig-Weg 3, 37077 G\"ottingen, Germany
\and{Earth Life Science Institute, Tokyo Institute of Technology, 2-12-1-IE-1 Ookayama, Meguro-ku, Tokyo 152-8550, Japan}
\and{Department of Physical Sciences, The Open University, Walton Hall, Milton Keynes, MK7\,6AA, UK }
\and{University of Bremen, Institute for Applied and Physical Chemistry, Leobener Str.~5, 28359 Bremen, Germany}
\and{German Aerospace Center, DLR, 51147 Cologne, Germany }
\and{Leiden Observatory, PO Box 9513, 2300 RA Leiden, The Netherlands }
\and{Space Policy Institute, George Washington University, 20052 Washington DC, USA }
\and{CREEST, Universities Space Research Association appointed at NASA Goddard Space Flight Center, Greenbelt, MD  20771, USA }
\and{PIIM, UMR7345, Avenue Normandie-Niemen, 13013 Marseille, France}
\and{Space Technology Ireland, Ltd., Maynooth, Co. Kildare, Ireland }
\and{LISA, UMR CNRS 7583, Universit{\'e} Paris-Est Cr{\'e}teil \& Universit{\'e} Paris-Diderot, 
94000  Cr{\'e}teil, France }
\and{Univ. Versailles St-Quentin; Sorbonne Univ., UPMC Univ. Paris 06; CNRS/INSU; LATMOS-IPSL, 4 Place Jussieu 75005 Paris }
}
\date{Accepted XXX. Received YYY; in original form ZZZ}

\abstract
{

{\sf \em Context.} The Rosetta lander Philae successfully landed on the nucleus of comet 67P/Churyumov-Gerasimenko on 12 November 2014.
Philae is equipped with two gas analyzers: The Cometary Sampling and Composition experiment (COSAC)   
and the gas chromatograph and mass spectrometer Ptolemy. 

{\sf \em Aims.}
 COSAC is designed for in situ analysis of organic molecules on 67P while Ptolemy is 
optimised to measure ratios of stable isotopes.

{\sf \em Methods.}
On 12 to 14~November 2014
both instruments measured the organic composition of the comet nucleus material through seven 
measurements in sniffing mode during Philae's hopping and at its final 
landing site Abydos.  
We  compare the temporal evolution of intensities of several ion species identified by both mass spectrometers.
 For COSAC this is the first analysis of the temporal behaviour of the measured ion species.

{\sf \em Results.}
All ion species showed the highest intensities in the first spectra measured by both instruments 
about 20 to 30~minutes after
Philae's first touchdown at Agilkia, and a decay during the six consecutive measurements at Abydos. 
Both instruments measured a nearly identical decay of the water peak ($m/z$ 18), and also CO 
($m/z$ 28) behaved similarly. 
In the COSAC measurements the peak at $m/z$ 44 decays much slower than all the other ion species, including 
the water peak.  In particular, the $m/z$ 44 peak decays much slower in the COSAC measurements
than in the Ptolemy data. This supports our earlier interpretation that COSAC 
for the first time analyzed a  regolith sample from a cometary nucleus in situ, while Ptolemy
measured cometary gas from the ambient coma. The $m/z$ 44 peak measured by COSAC  
was  likely dominated by  organic species, whereas the peak measured by Ptolemy was 
interpreted to be mostly due to $\mathrm{CO_2}$. Ion species heavier than $m/z$ 30 tend to decay somewhat slower in the
COSAC measurements than in the Ptolemy data, which may be related to differences in the
exhaust designs between both instruments. 
}

\keywords{
Astrochemistry, 
Comets individual: 67P/Churyumov-Gerasimenko, 
Comets: general, 
Astrobiology, 
Space vehicles: instruments}

\titlerunning{Decay of COSAC and Ptolemy mass spectra}

\maketitle

\bibliographystyle{aa}




\section{Introduction}

\label{sec_introduction}

On 12 November 2014 Philae successfully landed on its target comet 67P/Churyumov-Gerasimenko 
(hereafter 67P) at a heliocentric distance of 2.99 AU \citep{biele2015}. Philae carries two evolved gas analyzers 
on board: The Cometary Sampling and Composition (COSAC) experiment designed for in situ analysis
of organic molecules on 67P \citep{goesmann2007} and the gas chromatograph/mass spectrometer Ptolemy
optimised to measure ratios of stable isotopes \citep{wright2007}. 

\begin{figure*}
   \centering
  \hspace{-9mm}
\includegraphics[width=0.52\textwidth]{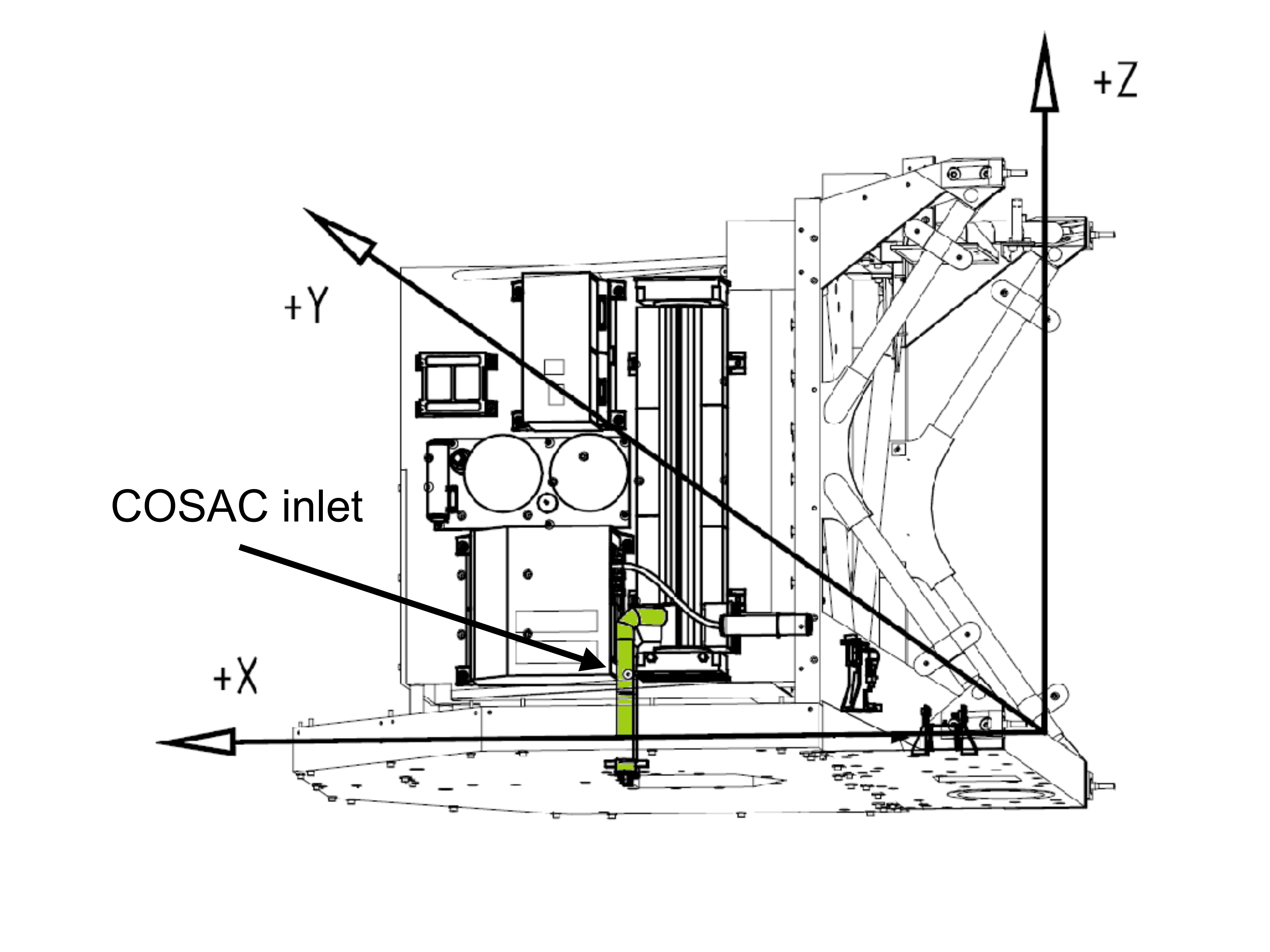}
\hspace{-8mm}
\includegraphics[width=0.52\textwidth]{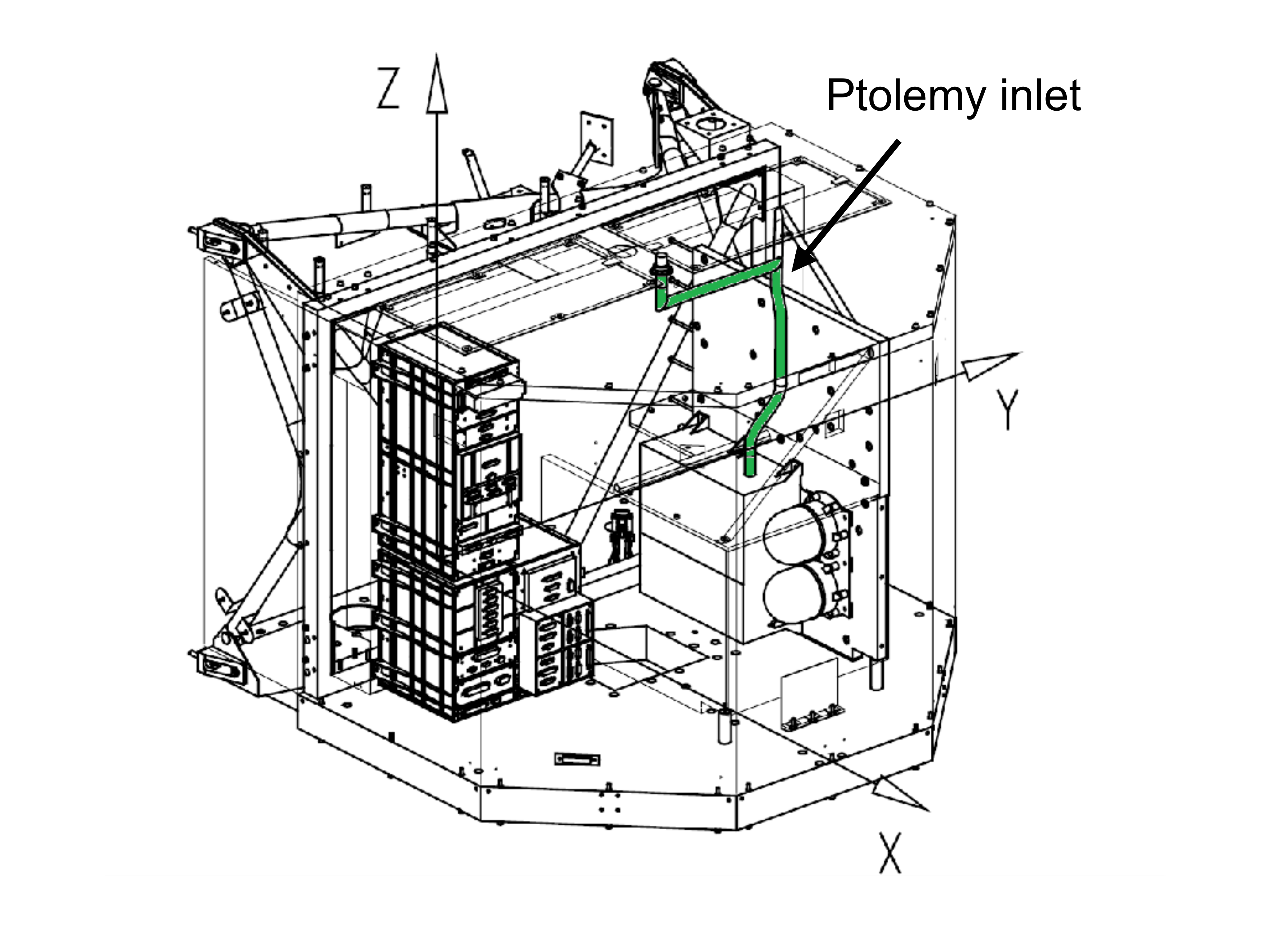}
\caption[]{
Three-dimensional drawings of Philae. The vent tubes (gas inlets) for COSAC ({\em left}) and Ptolemy ({\em right}) are indicated. The COSAC vent tubes have an inner diameter of 19 mm and a length of 145~mm and 165~mm, respectively. The 
siphon-shaped Ptolemy tube has an inner diameter of 12~mm and is approximately 500~mm long. 
}
\label{fig_phil}
\end{figure*}

Both instruments were successfully operated from 12 to 14  November 2014 during Philae's hopping and at its final 
landing site Abydos (during the so-called First Science Sequence of Philae). In this time interval
COSAC performed seven measurements in so-called sniffing mode (this term refers to a direct 
measurement of the ambient volatiles released from the comet, without prior gas chromatographic separation). 
Results from the first of these measurements obtained about 25 minutes
after the first touchdown at Agilkia were published by \citet{goesmann2015}, while the remaining 
six COSAC spectra obtained at Abydos were not previously described in detail. From the circumstances of the
first touchdown it is believed that COSAC collected with its twin
vent tubes -- located at the bottom of Philae pointing toward the cometary surface (Figure~\ref{fig_phil}, left panel)
-- a tiny amount of the cometary 
 regolith that was raised in a 
dust cloud during touchdown  and later sublimed inside the instrument \citep{goesmann2015}. A dust cloud was 
indeed recorded by the Rosetta navigation
camera close to Agilkia a few minutes after touchdown, approximately $\mathrm{0.4\,m^3}$ 
of material was excavated \citep{biele2015}.

Ptolemy measurements were taken 12 to 15 minutes before each of the seven COSAC measurements. 
Results were published by \citet{wright2015} and \citet{morse2015}. In the 
latter paper the authors consider the temporal evolution of three abundant cometary species: 
$\mathrm{H_2O}$, CO and $\mathrm{CO_2}$. Ptolemy likely collected ambient gas, which during the first measurement consisted of the ambient coma as well as additional gas from the recently disturbed regolith and dust cloud, because Ptolemy's vent tubes are located at the top of Philae and were pointing away from the cometary surface at the time of the measurements (Figure~\ref{fig_phil}, right panel).


In this paper we evaluate the time evolution 
of several ion species that were identified in the COSAC and Ptolemy spectra, with a particular emphasis on the peaks at a mass per unit charge
 $m/z$ 18, 28 and 44 ($\mathrm{H_2O}$, CO and partially $\mathrm{CO_2}$).

\section{Analysis}

\label{sec_analysis}

\begin{table*}
\tiny
   \centering
      \caption{Measurement times (Day and UTC), total ion counts (TIC) and peak intensities for $m/z$ species extracted from 
      the COSAC  spectra. }
   \begin{tabular}{@{} ccccccccccccccccccc @{}} 
   \hline    
   \hline
Meas.     &    Day     &   UTC &   TIC & \multicolumn{14}{c}{$m/z$} \\
Number    &   2014     & [hh:mm] &       & 15 & 16 &  17 &   18 & 27 & 28 & 29 & 30 & 31 & 43 & 44 & 45 & 57 & 59 & 61 \\
\hline
      0       & 16 Oct & 14:30 &  373  &  1 &  5 &   2 &   19 &  2 & 23 & 4  &  2 &  3 &  2 & 23 &  0 &  0 &  2 &  0  \\
      1       & 12 Nov & 16:02 & 19242 & 141& 59 & 536 & 2521 & 70 & 137& 172& 62 & 77 & 112& 121&114 & 95 & 65 & 30 \\
      2       & 13 Nov & 06:48 & 5161  &  10& 25 & 178 & 797  & 12 & 43 & 22 & 6  & 9  & 16 & 93 & 4  & 5  & 5  & -3 \\
      3       & 13 Nov & 08:50 & 2915  &  15& 20 & 91  & 393  &  6 &  35& 12 & 7  & 0  & 10 & 85 & 0  & 3  & -6 & -4 \\
      4       & 13 Nov & 10:52 & 1962  &   4&  0 & 56  &  210 &  1 & 23 & 16 & 11 & 8  &  2 &  74& 0  &  0 & -1 & -1  \\
      5       & 13 Nov & 12:54 & 1651  &   6& 10 & 47  & 188  &  8 & 28 & 16 & 3  & -2 &  9 & 60 & 1  & 1  & -1 &  4  \\
      6       & 14 Nov & 02:59 & 798   &   6&  9  &11  & 56   &  7 & 30 &  4 & 0  & 2  &  2 & 40 & 0  & 0  & 0  &  0  \\
      7       & 14 Nov & 12:44 & 728   &   0 & 6  & 11 & 63   &  1 & 20 & 2  & 0  & 1  &  8 & 46 & 1  & 2  &  0 &  2 \\
      \hline
      &  &  &    &    &   &  &    &   &  &   &   &   &   &  &   &   &   &   \\
         \end{tabular}
         
        {\bf Notes.}  The noise is about 10 ion counts per integer $m/z$ bin.
   \label{tab_1}
\end{table*}

\begin{table*}
   \centering
      \caption{Measurement times (Day and UTC), total ion counts (TIC) and peak intensities for $m/z$ species extracted from 
      the Ptolemy spectra. }
   \begin{tabular}{@{} ccccccccccccccc @{}} 
   \hline      
   \hline
Meas.     &    Day     &   UTC &   TIC & \multicolumn{11}{c}{$m/z$} \\
Number    &   2014     & [hh:mm] &       & 15 &     18 & 27 & 28  & 30  &  31 &  43 & 44  & 57 & 59 & 61 \\
\hline
      0       & 16 Oct & 17:23 &  171  &  2 &    112 &  2  &  5 &  2  &   0 &   0 & 10  &  3  &  1  & 0  \\
      1       & 12 Nov & 15:54 & 6721 & 150 &   2142 & 38 & 211& 45 & 314& 257& 575 & 414& 130& 79 \\
      2       & 13 Nov & 06:35 &  923  & 10 &    610 & 10 &  20&  1 &   7& 15 & 107 &   3&   3& 6 \\
      3       & 13 Nov & 08:37 &  455  &  6 &    304 &  2 &   9&  1 &  3 & 8  &  64 &   3&  1 & 1 \\
      4       & 13 Nov & 10:39 &  284  &  2 &    191 &  2 &   6&  1 &  1 & 7  &  38 &   1& 1  & 2   \\
      5       & 13 Nov & 12:41 &  117  &  8 &     79 &  1 &   2&  1 & 1  & 4  &  17 &   1& 1  & 1  \\
      6       & 14 Nov & 02:54 &   39  &  0 &     21 &  0 &   1&  0 & 0  & 4  &   6 &   0& 0  & 0  \\
      7       & 14 Nov & 12:36 &   54  &  0 &     22 &  0 &  2 &  0 & 1  & 2  &  23 &   0& 0  & 0  \\
      \hline
      &  &  &     &   &      &   &   &   &   &  &   &   &   &   \\
         \end{tabular}
         
         {\bf Notes.} Entries for $m/z$ 18 ($\mathrm{H_2O}$) are the sum of ion species at $m/z$ 16, 17, 18 and 19; those
      at $m/z$ 28 (CO) are the sum of $m/z$ 28 and 29 and those for $m/z$ 44 ($\mathrm{CO_2}$) are the sum of $m/z$ 44 and 45 \citep{morse2015}.
   \label{tab_1a}
\end{table*}

Philae reached the comet surface and made its first touchdown on 12 November 2014, 15:34:04~UTC at the 
Agilkia landing site. Instead of a nominal landing, Philae bounced and continued its journey across the 
surface of 67P and 
finally came to rest at Abydos at 17:31:17~UTC, about two hours after its first touchdown. The first 
COSAC spectrum was taken about 25 minutes after the first touchdown during Philae's hoppings, and 
the following six spectra were collected at Abydos on 13 and 14 November 2014 (Table~\ref{tab_1}). An 
additional COSAC spectrum that was taken on 16 October 2014 at a distance of about 10~km from the nucleus 
we take as a reference for the instrumental background  (measurement \#0 in Table~\ref{tab_1}). 
The instrumental settings 
were identical for all  measurements so that we can directly compare these data.

Our analysis is based on peak lists generated from each spectrum.  The peak lists contain the
total number of ions  counted in predefined mass bins
centred close to the integer $m/z$ of the respective peak, 
after subtraction of background noise determined in separate bins adjacent to the peak. 
Table~\ref{tab_1} lists the total number of ions for the most intense peaks in the COSAC
spectra. Details about the algorithm for generating the peak lists can be found in 
\citet[][Supplementary Material]{goesmann2015}.

\begin{figure*}
\vspace{-6mm}
   \centering
   \hspace{-9mm}
      \vspace{-0.5cm}
\includegraphics[width=0.51\textwidth]{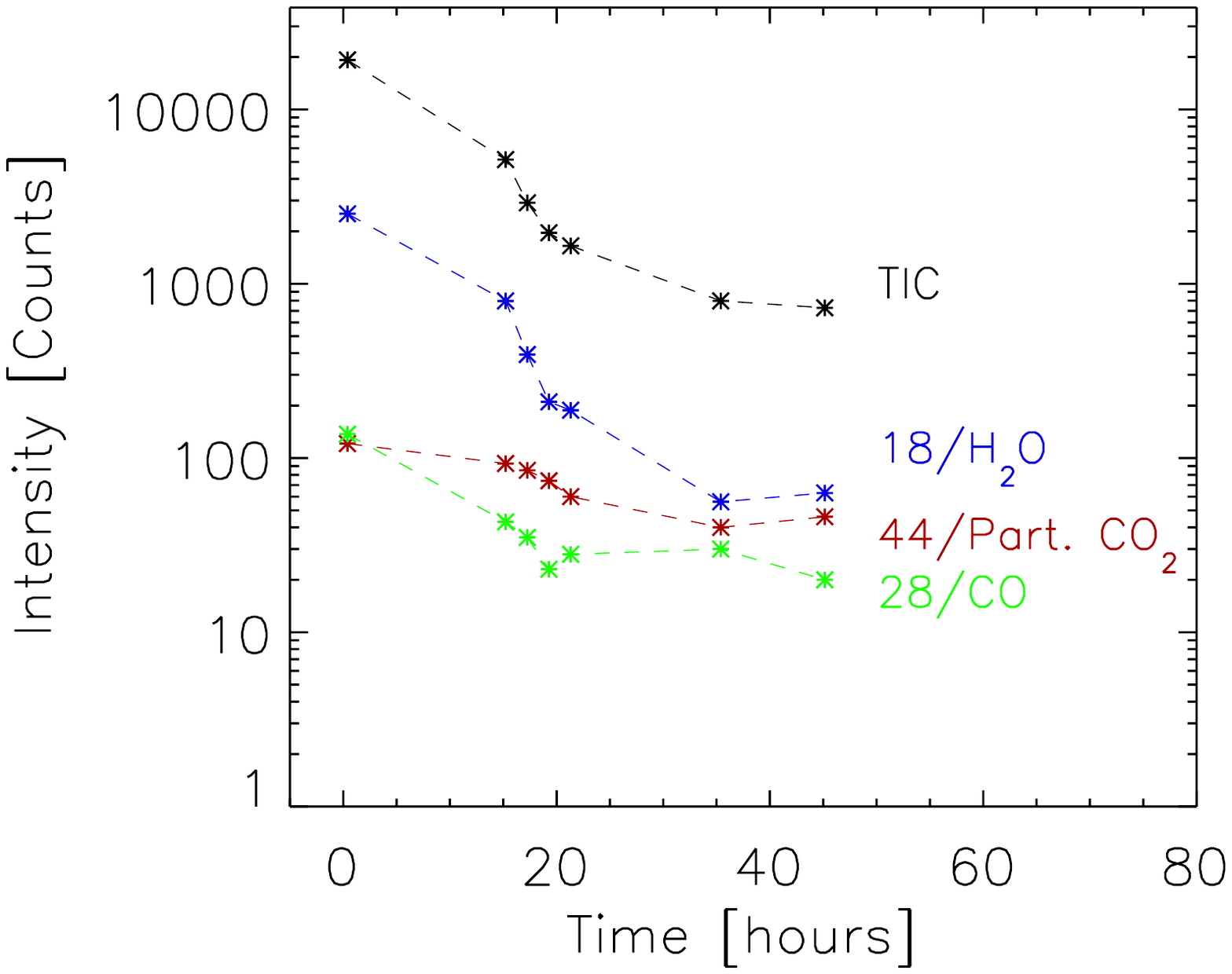} 
\hspace{-16mm}
\includegraphics[width=0.51\textwidth]{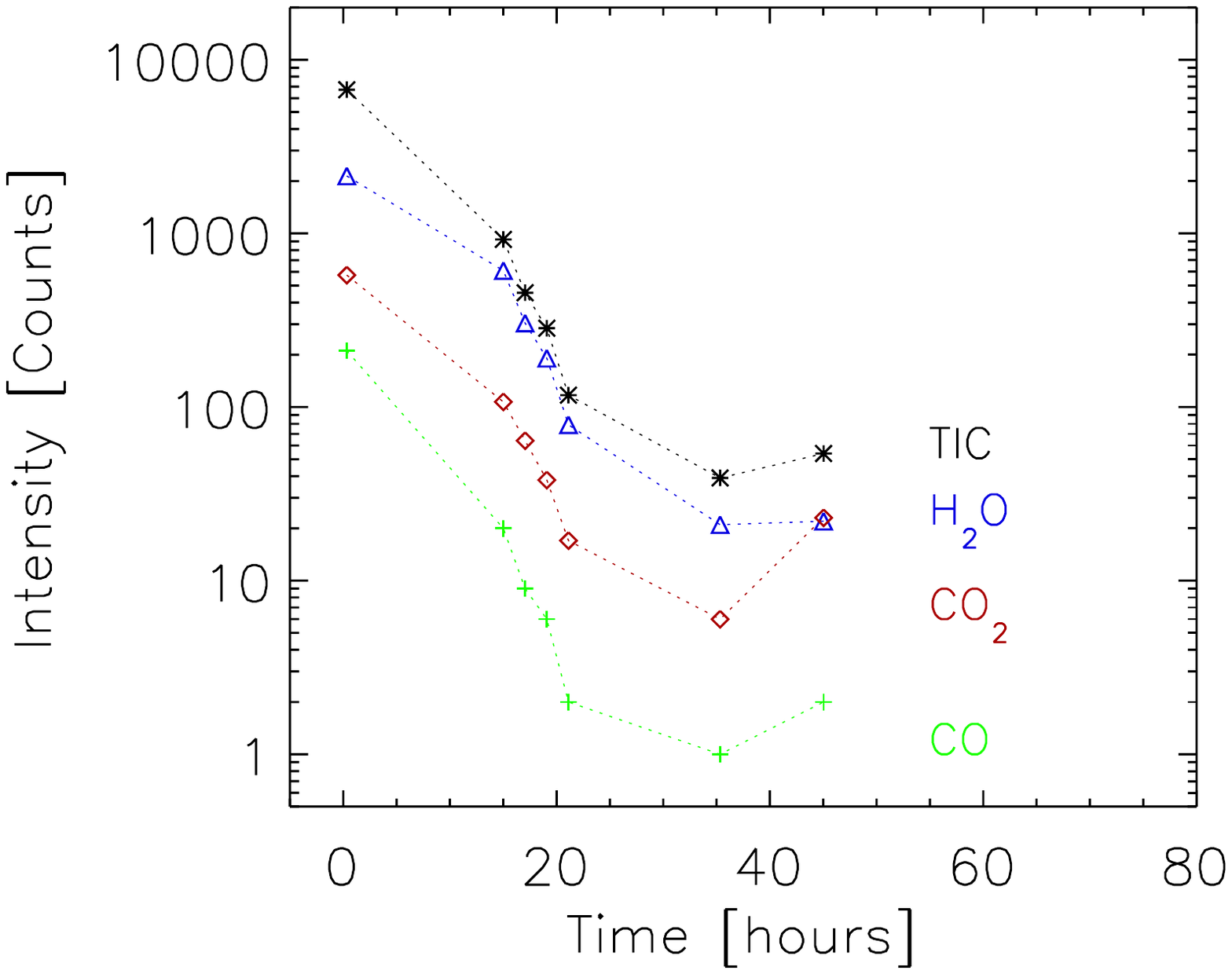}
      \vspace{-5.5cm}
\caption[]{
{\em Left:} Temporal behaviour of ion species with $m/z$ 18 ($\mathrm{H_2O}$), 28 (CO), 44 (partially $\mathrm{CO_2}$) and total ion counts (TIC), 
respectively, derived from the seven COSAC spectra (raw counts, not corrected for instrumental background). 
{\em Right:} Ptolemy data taken from \citet[][their Tab.~3, corrected for hydrogen abstraction, a feature of the ion trap employed in Ptolemy]{morse2015}. The time axis  is set to zero at Philae's touchdown on 12 November 2014, 15:34:04~UTC 
(see text for details). 
}
\label{fig_1}
\end{figure*}

Figure~\ref{fig_1} (left) shows peak intensities derived from the COSAC  peak lists.  
The noise level is approximately 10 ion counts per integer $m/z$ bin, and only in the first 
spectrum do many ion species show a much higher count rate significantly above the noise level. 
Intensities extracted from the first five spectra for ion species $m/z$ 18, 28, 44 and for the 
total ion counts (TIC) show an approximate exponential drop during the initial 22 hours after 
Philae's first touchdown. Peak intensities for the other species are very 
much influenced by noise (see Table~\ref{tab_1}).  The peak 
intensities for all ion 
species tend to level off approximately 20 hours after touchdown,  indicating that they 
become dominated by noise. We therefore ignore the last two COSAC spectra in our analysis.  

\begin{figure*}
\vspace{-8mm}
   \centering
   \hspace{-9mm}
   \vspace{-0.5cm}
\includegraphics[width=0.51\textwidth]{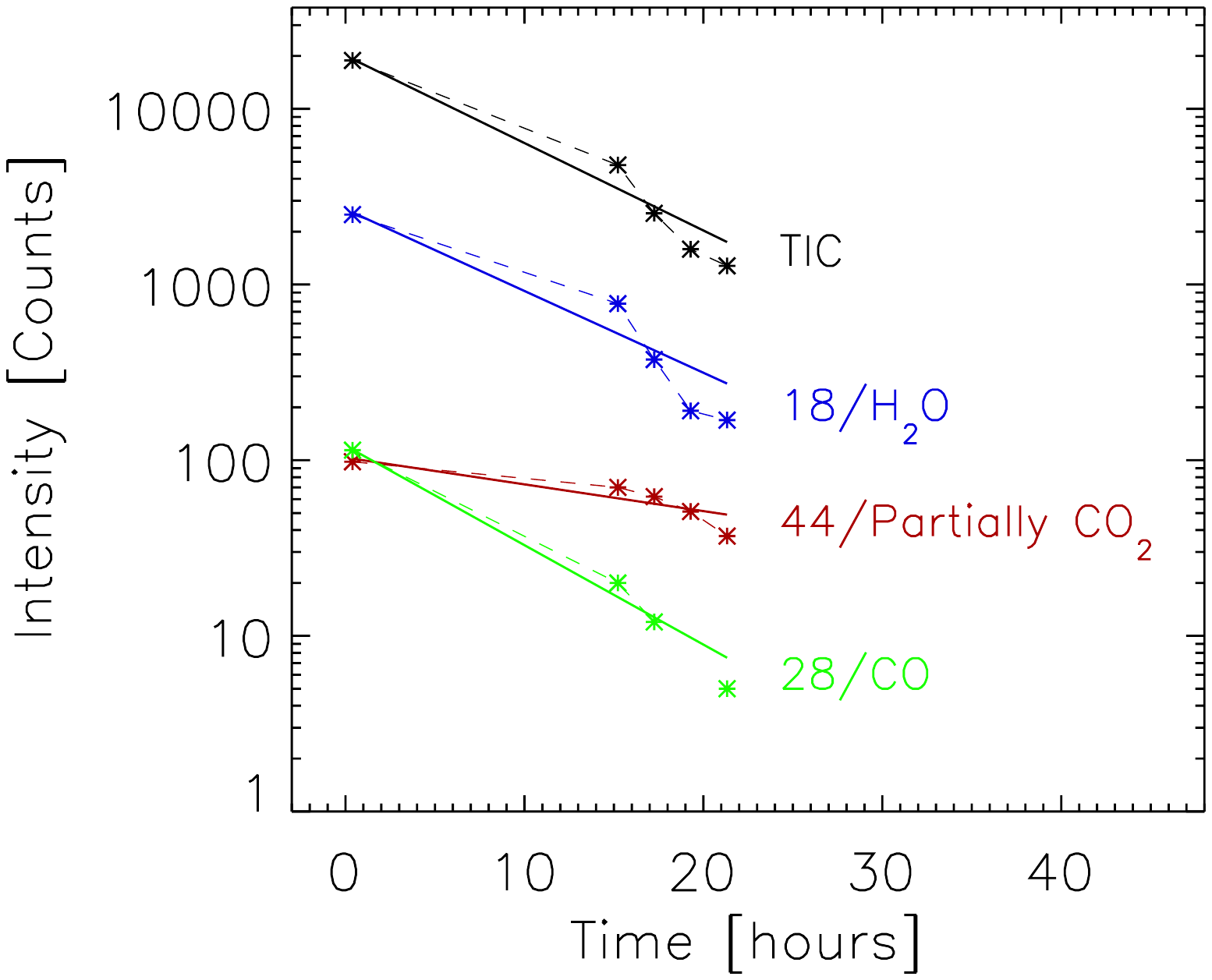}
\hspace{-16mm}
\includegraphics[width=0.51\textwidth]{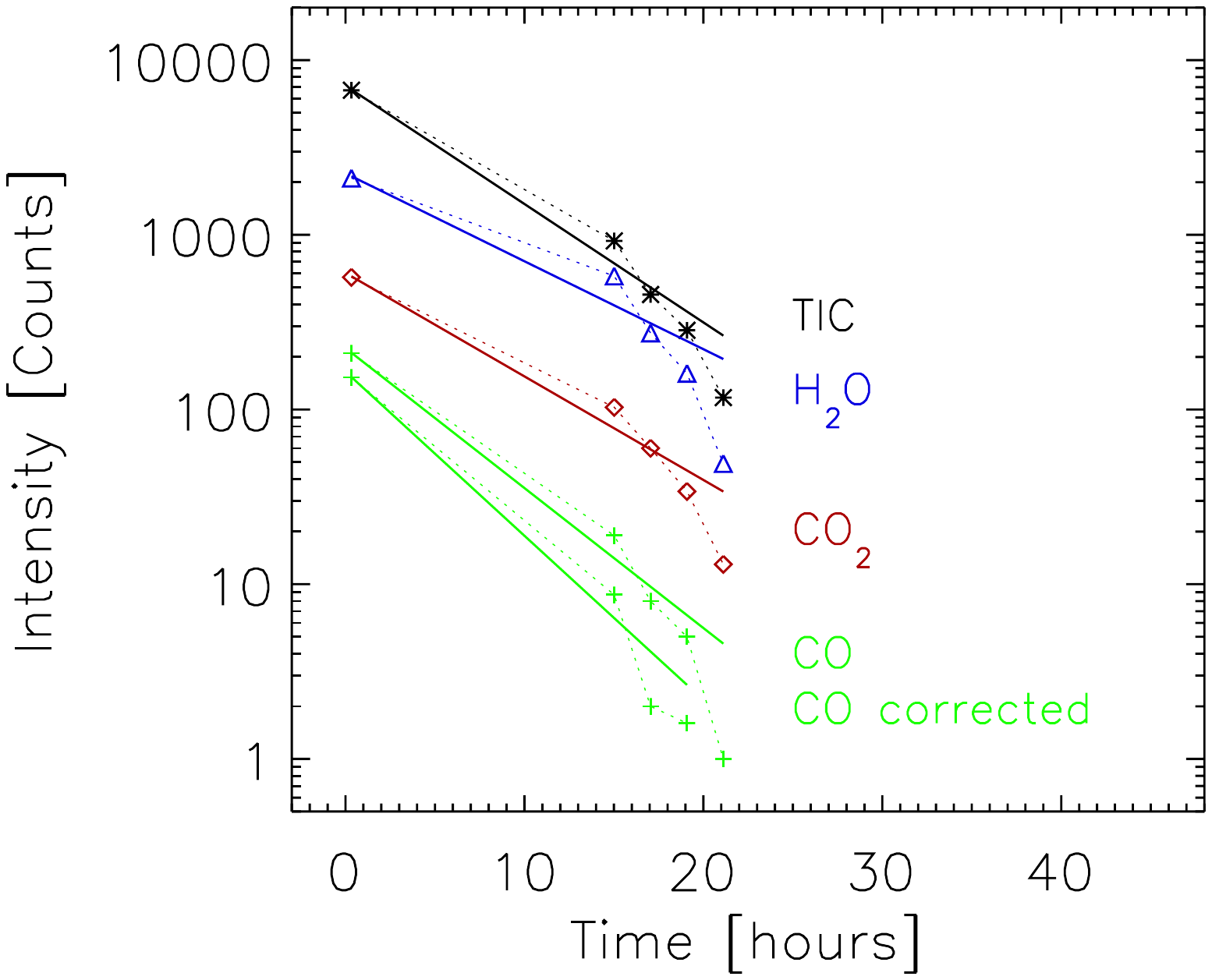}
   \vspace{-5.3cm}
\vspace{-2mm}
\caption[]{
{\em Left:} COSAC measurements \#1 to 5 (Table~\ref{tab_1}) corrected for instrumental backgound. Spectrum \#0 was 
used for background correction.  The noise is about 10 ion counts per integer $m/z$ bin.
{\em Right:} Same for Ptolemy. Corrections for instrumental background and  $\mathrm{CO_2}$ fragmentation were applied
(see text for details). Exponential slopes were fitted to the data (Table~\ref{tab_2}). As for Figure~\ref{fig_1}, the time axis is set to zero at Philae's touchdown.
}
\label{fig_2}
\end{figure*}

Figure~\ref{fig_1} (right panel) shows Ptolemy data for $\mathrm{H_2O}$, CO and $\mathrm{CO_2}$ measured between 
12 and 14 November 2014 \citep{morse2015}. It should be noted that the ion counts given for these 
three species are summed ion counts for up to
four $m/z$ bins to account for hydrogen abstraction (i.e. the addition or subtraction of one or more 
protons to the ion species under consideration): $m/z$ 16, 17, 18 and 19 to 
derive the intensities of $\mathrm{H_2O}$, $m/z$ 28 and 29 for CO and $m/z$ 44 and 
45 to get the intensity of $\mathrm{CO_2}$. The analysis of the Ptolemy data was described in detail by
\citet{morse2015}. 
The peak intensities for $\mathrm{H_2O}$, CO and $\mathrm{CO_2}$ measured with Ptolemy
level off 
after approximately 20~hours and display a  flattening similar to that found in the COSAC 
measurements. Therefore, we also ignore these two final Ptolemy spectra. 

We assumed that   
comet material entered the COSAC vent tubes (Figure~\ref{fig_phil}, left panel) which -- during Philae's first touchdown at Agilkia -- 
pointed downwards, i.e. towards the cometary nucleus \citep{goesmann2015}.  Once inside COSAC the 
material was heated to $\mathrm{\sim 280~K}$  so that
gases  could evolve from the grains. In the ideal case of a closed system and with enough material available, the partial pressure in the system would have increased with time up to the relevant vapor pressure at the tube temperature.
The real system, however, is not closed and gas could escape from the tubes. Therefore,  we assume an 
approximately exponential pressure drop in the tubes.

No thermal sensors are located on the vent tubes of COSAC, therefore the temperature can only be 
derived indirectly based on the temperatures of the closest thermal sensors on board Philae and on 
simulations with a thermal mathematical model. The analysis based on 
the temperature recorded during Philae's operation 
at 67P by two COSAC sensors  at the electronic box and on the gas chromatograph and the Philae baseplate sensor  showed a temperature variation from the first touchdown to the end of Philae's operation at Abydos.  The simulations showed that
the temperature at the COSAC mass spectrometer likely dropped from $\mathrm{\sim 280~K}$ at first touchdown 
to $\mathrm{\sim 260~K}$ \citep{cozzoni2016} at the time when measurement \#5 was performed, and a similar drop can be expected for the exhaust tubes as 
well. The Ptolemy internal temperature sensor measured a drop by only 4~K from 
273~K to 269~K between measurement \#1 and \#5 \citep[][their Tab.~1]{morse2015}.

Even though the temperature of COSAC dropped by about 20~K during the analysis period reported here, it remained much higher than the cometary environment and should have only marginally contributed toward species decay in the instrument. We therefore make the simplifying assumption
that the temperature remained constant. We believe this is a 
reasonable assumption because also at about 260~K, the gas pressure in the system remained far 
below the vapor pressure of the three species considered  (otherwise our mass spectra would have been 
saturated). Then 
 -- to a first order approximation --
the pressure of each individual species should follow an exponential decay curve. 
For Ptolemy we made a similar assumption with the exception that ambient coma gas entered the vent 
tubes of Ptolemy, which pointed away from the comet surface. 

In Figure~\ref{fig_2} we show the temporal behaviour of $\mathrm{H_2O}$, CO and 
$\mathrm{CO_2}$ in the COSAC and Ptolemy spectra during the initial 
22~hours after touchdown at Agilkia. 
(measurement \#1 to \#5 in Tables~\ref{tab_1} and \ref{tab_1a}).  For both instruments in measurement \#1 all peaks 
were above the background noise level while in the later spectra many peaks, in particular all peaks above 
$m/z$ 45, had dropped to the background level. 
For COSAC we subtracted the peak intensities derived from this reference spectrum from 
the other measurements in order to correct for instrumental background.  Note that the 
background of $m/z$ 18 was quite low and thus the correction is not significant; 
only the results for $m/z$ 28 are affected by the background correction. 

It should be noted that in our analysis of the COSAC spectra we also consider peaks for ion species with a rather low count 
rate. This
applies in particular to peaks at $m/z$ 30, 31, 45, 57 and 59. In these cases, only the COSAC spectrum \#1 showed 
 strong intensities well above the noise level, which
is estimated at about 10 counts per integer $m/z$ bin. In the other spectra measured 15 to 22 hours later 
these peaks are very weak and, thus, the slopes we derive here are likely
 lower limits for the true slopes. Hence, the peaks for these species likely dropped even faster than implied by our slope
 fitting. 
 
The Ptolemy data for $\mathrm{H_2O}$, CO and 
$\mathrm{CO_2}$ were taken from \citet[][their Tab.~3]{morse2015}. 
 To the CO peak for Ptolemy we  applied a correction for the contribution of $\mathrm{CO_2}$
fragments to the CO peak. We  assumed a ratio of 10/1 based on the NIST database
\citep{stein2015,morse2015}. The corrected and uncorrected data are compared in Figure~\ref{fig_2}. 

We also applied a correction for instrumental background to the 
Ptolemy data. We assumed 30, 2 and 4 counts for
$\mathrm{H_2O}$, CO and $\mathrm{CO_2}$, respectively, measured beyond 30~km from the nucleus \citep{morse2015}. 
The effect of this correction is very small, changing the fitted slope only in the third digit after the 
decimal point. We assumed a zero background for all the other species. The correction for instrumental 
background in the Ptolemy data is thus totally negligible in our analysis. 
 

\section{Results}

\label{sec_discussion}

\begin{table}
   \centering
      \caption{Slopes of exponential functions fitted to peak intensities derived from the COSAC and Ptolemy spectra. }
   \begin{tabular}{@{} cccc @{}} 
   \hline      
   \hline
$m/z$ &       \multicolumn{2}{c}{Slope} \\ 
      &    COSAC            & Ptolemy \\
\hline
TIC   &   $-0.12 \pm 0.01$  & $-0.16 \pm 0.01$ \\
15    &   $-0.16 \pm 0.02$ & $-0.17 \pm 0.02$ \\
16    &   $-0.08 \pm 0.01$ &      --      \\
17    &   $-0.10 \pm 0.01$ &      --      \\
18    &   $-0.11 \pm 0.01$ & $-0.12 \pm 0.01$ \\
27    &   $-0.13 \pm 0.01$ & $-0.13 \pm 0.03$ \\
28    &  $-0.13  \pm 0.05$ & $-0.18 \pm 0.02$ \\
 29   &  $-0.14  \pm 0.01$ &       -- \\
 30   &  $-0.13  \pm 0.02$ & $-0.20 \pm 0.05$ \\
 31   &  $-0.15  \pm 0.02$ & $-0.27 \pm 0.04$ \\
 43   &  $-0.14  \pm 0.01$ & $-0.20 \pm 0.02$ \\
 44   &  $-0.04  \pm 0.01$ & $-0.14 \pm 0.01$ \\
 45   &  $-0.23  \pm 0.04$ &       --    \\
 57   &  $-0.20  \pm 0.03$ & $-0.31 \pm 0.04$ \\  
  59   &  $-0.21  \pm 0.04$ & $-0.27 \pm 0.05$ \\ 
  61   &  $-0.10 \pm  0.05$ & $-0.20 \pm 0.04$ \\
      \hline
      &    & \\
         \end{tabular}
         
         {\bf Notes.} A 
      $\sqrt{n}$ weighting of the individual data points was used for the fits, $n$ being the intensity (counts) of the respective peak. The $\pm$ uncertainties given are  
$1 \sigma$ uncertainties for the slopes of the fit curves.
   \label{tab_2}
\end{table}

\begin{figure*}
   \centering
   \vspace{-0.3cm}
\includegraphics[width=0.7\textwidth]{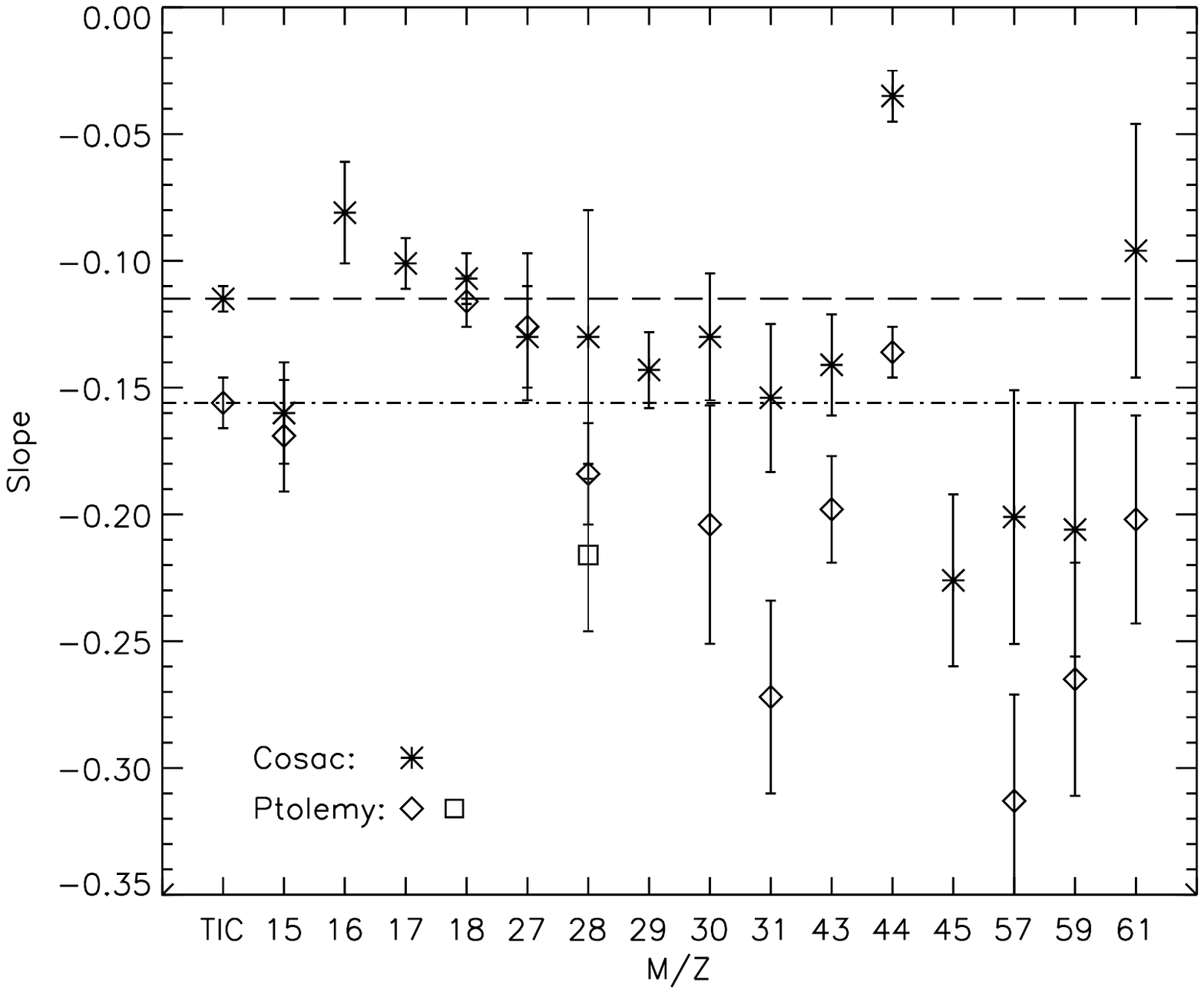}
\vspace{-8.1cm}
\caption[]{
Slopes of fits for the individual ion species investigated; asterisks: COSAC (data listed in Table~\ref{tab_2}), diamonds: Ptolemy, square: Ptolemy $m/z$ 28 corrected for $\mathrm{CO_2}$ fragmentation. The error bars display $1\sigma$ uncertainties for the slopes derived from fitting exponentials to the data, taking into account uncertainties in instrumental background. The horizontal lines indicate
the values for the total ion counts (TIC) for COSAC (long dashes) and Ptolemy (dash-dots). 
}
\label{fig_3}
\end{figure*}

In Figure~\ref{fig_3} we compare the slopes derived from fitting exponentials to the count rates of 
the individual peaks in the COSAC and Ptolemy spectra.  For the fit each data point was weighted with 
the square root of the total number of ion counts in the peak. This assures that peaks with a
higher statistical significance get a higher weight in the  fit. 
Figure~\ref{fig_3} shows that the temporal behaviour differs 
between individual ion species.

\subsection{COSAC}

In the COSAC spectra the peak at $m/z$ 44 exhibits the slowest drop of all peaks considered in our 
analysis. It decays  
significantly slower than, for example, the water peak at $m/z$ 18. For both 
COSAC and Ptolemy  the peak at 
$m/z$ 18 is by far the strongest in the spectra (Tab.~\ref{tab_1} and \ref{tab_1a}). Furthermore, in 
the COSAC spectra the slope for the total 
ion counts (TIC) is indistinguishable within the measurement accuracy from the slope of the 
$\mathrm{H_2O}$ peak (Tab.~\ref{tab_2}) . 

For COSAC the slopes 
tend to fall into two or, possibly, three categories: 
\begin{itemize}
\item[A)] Most of the species show a trend very similar to the water peak. This is particularly obvious for $m/z$ 17, 27, 28 and 30. 
\item[B)] $m/z$ 44 clearly shows a flatter slope than all other species.
Note that \citet{goesmann2015} assigned the $m/z$ 44 peak to a large extent to species 
other than $\mathrm{CO_2}$ (see Section~\ref{sec_discussion} for details).
\item[C)] $m/z$ 15, 45, 57 and 59 show a somewhat steeper decrease than water. However, given the low number of counts, all 
 of these species may also be compatible with the water peak. 
\end{itemize}

\subsection{Ptolemy}

In Figure~\ref{fig_3} the entries for $m/z$ 18, 28 and 44 for Ptolemy are the summed up values from 
several other peaks: The groups defined by \citet{morse2015} -- water, CO, 
$\mathrm{CO_2}$, TIC and 'other' --
were defined to correct for the effects of hydrogen abstraction, a feature of the ion trap 
(see also Section~\ref{sec_analysis}).  We can therefore directly compare the slopes for these species 
derived from COSAC and Ptolemy. 
For ion species at other $m/z$ values, hydrogen abstraction can shift the peak location depending on 
the affected molecule and without a detailed knowledge of the underlying molecule, no correction for hydrogen abstraction can be applied. 

The results from Ptolemy are generally similar to the COSAC results, except for $m/z$ 44 (mostly $\mathrm{CO_2}$ in the 
case of Ptolemy). A variation in the decay of the peak intensities is also evident in the Ptolemy spectra. 
The $m/z$ 18  peak shows striking agreement with the value measured by COSAC. For Ptolemy, the water peak decays somewhat slower than do the total ion counts. 

Up to $m/z$ 30 the peaks in the Ptolemy spectra show a similar trend to those found for COSAC, while heavier species tend to decay somewhat faster than for COSAC. The most significant difference occurs for $m/z$ 44. For Ptolemy the decay of this peak is close to the value measured for water while for COSAC, this peak decays much more slowly.  It supports our earlier 
interpretation that Ptolemy and COSAC measured different samples of cometary material (see Section~\ref{sec_discussion}).

\section{Discussion}

\label{sec_discussion}

The interpretation of the COSAC data put forward by \citet{goesmann2015} 
is in conformity with data interpretations from other scientific payloads made both aboard the Philae Lander 
and the Rosetta Orbiter. For instance, data from the Surface Electrical Sounding and Acoustic Monitoring Experiment (SESAME) on 
Philae  and Philae's landing gear indicated that Agilkia apparently has a soft and granular surface ($\sim 20\,\mathrm{cm}$ in thickness) on top of a more rigid, icy subsurface  \citep{roll2016a,roll2016b}. 
In contrast, Abydos was found to  consist mostly of rigid, icy material \citep{biele2015,spohn2015}. 

The Visual Infrared and Thermal Imaging Spectrometer (VIRTIS) on board Rosetta revealed that solar insolation causes sublimation of $\mathrm{H_2O}$ ice, mostly from the uppermost surface layers. When in shadow or on the night-side a temperature inversion is present between the now colder surface and a subsurface that remains warmer for a longer duration. These observed diurnal variations probably trigger the sublimation-condensation cycle on the nucleus of 67P \citep[][see also below]{desanctis2015}.

On 12 November, when Philae impacted on the solar-illuminated Agilkia, it excavated $\sim 0.4\,\mathrm{m^3}$ 
of regolith from 67P's top surface layer \citep{biele2015}. \citet{desanctis2015}  discussed 
in detail a temperature gradient that exists between the top layer of the surface, which is 
reported to have a temperature between about 180~K and 230~K during daytime whereas the temperature of the 
subsurface, at a depth of 4~cm is, even during  daytime,  only about 130~K. Therefore, we infer
that Philae's 
excavation would have suddenly exposed the colder subsurface to the higher temperatures of the 
solar insolated surface and that this process would have rapidly caused the sublimation of volatile species from dust grains
present in the subsurface. Further, these volatilized species entered 
the COSAC and Ptolemy mass spectrometers. In the case of COSAC, the measurement was  likely dominated by gases which 
vaporized from grains after they entered the exhaust tubes and were warmed up later by the lander 
interior rather than sunlight. COSAC 
was at a much higher temperature ($\sim 260\,\mathrm{K}$ to 280~K) as measured by the thermal 
sensors on the COSAC electronic box and on the gas chromatograph \citep{goesmann2015,cozzoni2016}.

COSAC measured six mass spectra at the final landing site, Abydos. However, unlike in the first mass 
spectrum, these spectra did not generate peaks that could be attributed to organic species. 
Abydos is a geologically and thermally distinct site when compared to Agilkia. Abydos was almost
permanently in shadow with low diurnal surface temperatures ranging from 90~K to 130~K and 
has a hard sintered surface with limited regolith \citep{spohn2015}. The lower temperature  likely
prohibited sublimation of volatile organic species which thus could not diffuse into the COSAC 
mass spectrometer during all later six sniffing mode measurements.  Very low cometary activity  
close to Philae during the COSAC and Ptolemy measurements at Abydos was also indicated by measurements 
with the Dust Impact Monitor on board Philae which did not detect any dust impact during five hours 
of measurement time \citep{krueger2015b}.
 
We emphasize the distinct locations of COSAC's and Ptolemy's vent tubes 
that enabled their respective sniffing mode mass spectrometer operations. COSAC's vent 
tubes are located at the bottom plate of the lander and opened toward the surface 
of the nucleus. However, Ptolemy's vent tubes are located on top of the lander and opened  
toward the ''zenith`` (i.e. the cometary coma and open space;  cf. Figure~\ref{fig_phil}).
The COSAC tubes do not have a cover at their orifice, while the Ptolemy tube is
shielded by a table-like structure so as to prevent in-falling dust from entering directly.
It is  about 5~cm x 5~cm in size and located about 2~cm above the vent tube's orifice.

\citet{goesmann2015} during the course 
of their investigations had initially ascribed $m/z$ 44 in the first COSAC mass spectrum entirely to 
$\mathrm{CO_2}$; but doing so did not yield an appropriate mass fit to the other peaks especially 
those at $m/z$ 57, 58 and 59. They however did not rule out a minor contribution from $\mathrm{CO_2}$ 
to the $m/z$ 44 peak. The  most relevant species that they ascribed to this $m/z$ 44 peak and which 
generated a sensible fit with other peaks were acetaldehyde ($\mathrm{C_2H_4O}$, 
 accounting for approximately 56\% of the total peak intensity), 
 formamide ($\mathrm{HCONH_2}$, 21\%) 
and acetamide ($\mathrm{CH_3CONH_2}$, 23\%). 
 The interpretation of this fit to the first COSAC spectrum is presently under 
re-investigation (K. Altwegg et al., personal communication).

Contrary to COSAC's mass 
spectral interpretation, \citet{wright2015} concluded that $\mathrm{CO_2}$ was the major contributor to
the $m/z$ 44  and 45 peaks measured by Ptolemy.  Acetaldehyde, formamide and acetamide are comparatively 
less volatile than $\mathrm{CO_2}$
and could possibly be a constituent species of the subsurface cometary grains that probably entered 
the COSAC vent tubes immediately after impact excavation. Owing to the zenith directed 
position of Ptolemy's vent tubes and a siphon-shaped inlet (Figure~\ref{fig_phil}, right panel), also preventing grains from entering the system, 
its mass spectrometer could not have received organic-abundant 
grains from the impact-excavated subsurface directly. 
During the landing event, the coma composition would have been modified by gas vaporising from newly 
excavated dust exposed to solar illumination and contact with the lander (including the entrance to 
the Ptolemy vent tubes), thus explaining the high concentration of organics measured compared to orbiter measurements \citep{haessig2015}.
\citet{wright2015} therefore confirm measuring 
gaseous $\mathrm{CO_2}$ present in the nebulous coma atmosphere over the illuminated nucleus surface. The 
different slopes of the $m/z$ 44 peak as measured by both COSAC and Ptolemy indicate they have 
indeed analyzed different species, sampled via their distinctly positioned vent tubes, and 
originating from different environments on the comet.

 A requirement to support our interpretation  would be precise information about the phase transitions or vapour pressures of volatile organic species at extremely low temperatures and low-pressure conditions (typical of cometary environments). Phase transition studies of most organic species are derived using mathematical equations  such as
the Clausius-Clapeyron relation, Lee-Kesler 
method and Antoine equation. These and other similar equations usually generate appropriate vapour 
pressure values of chemical species existing under terrestrial conditions. Few empirical attempts to extrapolate 
the sublimation-condensation curves of volatile species $\mathrm{H_2O}$, CO and $\mathrm{CO_2}$ 
at extremely low pressures and temperatures were made in the past 
\citep{yamamoto1983,prialnik2005,huebner2006,fray2009}. However empirical extrapolations of vapour 
pressures at astrophysically relevant temperatures are not always accurate \citep{fray2009}. Even 
if accurate, most of these extrapolations are made assuming the species to be each in a pure ice configuration. 

A comet's surface material is heterogeneous and consists of complex refractory organic material, 
inorganic minerals, and a mixture of ices of several chemical species. For example, VIRTIS has 
demonstrated the 
presence of non-volatile refractory organic material on the surface that could be formed by ultraviolet 
or energetic particle irradiation of high-volatility carbon-containing ices like 
$\mathrm{CH_4}$, CO and $\mathrm{CO_2}$. So the precise vapour pressures of chemical species in 
such complex mixtures cannot be derived from the values of their pure ice counterparts. 
Determination of vapour pressures of organic chemical species existing in diverse astrophysical 
environments is a robust task that should be carried out before related astrophysics and astrochemical studies can be mounted.

 The situation is further complicated by the fact that species embedded as minor components in a 
matrix do not sublimate at their pure ice sublimation temperatures or a temperature related to their 
vapor pressure. Instead, they can be released at higher temperatures by so-called ``volcano'' 
desorption due to phase changes occurring in the ice between different amorphous structures or
from an amorphous to a crystalline phase, or by codesorption with water. Only very volatile 
species can desorb to a small part already at lower temperatures 
\citep{collings2004}. These desorption processes were studied in the laboratory with 
cometary analogue materials \citep{martin-domenech2014}. In the experiments the desorption occured
at temperatures below 180~K, i.e. at temperatures below the likely 
daytime temperature at the cometary surface (see above), rendering them rather unlikely in our case. 
At present, however, we cannot exclude that such desorption occurred in the cometary 
sample analyzed by COSAC.

%

\section{Summary}

\label{sec_conclusions}

We have investigated the temporal behaviour of various ion species in mass spectra measured with the mass 
spectrometers COSAC and Ptolemy on board the Rosetta lander Philae. Each instrument measured seven spectra, one during Philae's 
hopping about 20 to 30 minutes after touchdown at Agilkia and six consecutive spectra at Philae's 
final landing site Abydos. Results from the first COSAC measurements obtained about 25 minutes
after the first touchdown at Agilkia were published by \citet{goesmann2015}. It is the first 
in situ analysis of a nucleus  regolith sample from a comet. The remaining 
six COSAC spectra obtained at Abydos were not previously presented in detail.
The Ptolemy data were discussed by \citet{wright2015} and \citet{morse2015}. 

The ion intensities measured with both instruments are strongest in the spectra measured right
after Philae's touchdown at Agilkia and decreased later after Philae came to rest at Abydos. 
For most ion species, in particular water, the measured decay rates are in good agreement between 
both instruments. 
A striking difference is seen for $m/z$ 44: This peak drops significantly more slowly
in the COSAC spectra than in the Ptolemy spectra and, in the COSAC spectra, it exhibits 
the shallowest drop of all ion species investigated. 

The difference in the slopes of the $m/z$ 44 peak as measured by COSAC and Ptolemy indicates that both instruments
measured different ion species at this $m/z$. It  supports earlier results by \citet{goesmann2015} 
and \citet{wright2015},
indicating that COSAC measured for the first time cometary nucleus material from the Agilkia landing site
while Ptolemy measured the 
composition of the ambient gas, i.e. coma plus gas vaporised from disturbed dust. 
Unfortunately, the  limited operational lifetime of Philae
did not allow for further measurements and improved data.  


\section*{Acknowledgements}

This research was supported by 
the German Bundesministerium f\"ur Wirtschaft und Energie through Deutsches
Zentrum f\"ur Luft- und Raumfahrt e.V. (DLR, grant 50\,QP\,1302). Support by MPI f\"ur Sonnensystemforschung
is gratefully acknowledged as is the Programme de Development d'Exper\-iences Scientifiques administered by Enterprise Ireland (SMcKL).  FR and CS wish to thank CNES for its financial support to COSAC activities. CG gratefully acknowledges ELSI Origins Network (EON) which is supported 
by a grant from the John Templeton Foundation. We thank the Rosetta project at ESA, the Philae project 
at DLR and CNES and the Philae Lead Scientists for effective and successful mission operations. We would like to acknowledge the work of Helmut Rosenbauer and Colin Pillinger who are both unfortunately no longer with us. They envisaged the instruments and made them become reality so that we can harvest what they seeded.





\begin{thebibliography}{20}
\expandafter\ifx\csname natexlab\endcsname\relax\def\natexlab#1{#1}\fi

\bibitem[{{Biele} {et~al.}(2015){Biele}, {Ulamec}, {Maibaum}, {Roll}, {Witte},
  {Jurado}, {Mu\~noz}, {Arnold}, {Auster}, {Casas}, {Faber}, {Fantinati},
  {Finke}, {Fischer}, {Geurts}, {G\"uttler}, {Heinisch}, {Herique}, {Hviid},
  {Kargl}, {Knapmeyer}, {Knollenberg}, {Kofman}, {K\"omle}, {K\"uhrt},
  {Lommatsch}, {Mottola}, {Pardo de Santayana}, {Remetean}, {Scholten},
  {Seidensticker}, {Sierks}, \& {Spohn}}]{biele2015}
{Biele}, J., {Ulamec}, S., {Maibaum}, M., {et~al.} 2015, Science, 349, aaa9816

\bibitem[{{Collings} {et~al.}(2004){Collings}, {Anderson}, {Chen}, {Dever},
  {Viti}, {Williams}, \& {McCoustra}}]{collings2004}
{Collings}, M.~P., {Anderson}, M.~A., {Chen}, R., {et~al.} 2004, Monthly Notice
  of the Royal Astro. Soc., 354, 1133

\bibitem[{{Cozzoni}(2016)}]{cozzoni2016}
{Cozzoni}, B. 2016, {Temperature of COSAC Vent-pipe during FSS, Philae internal
  technical note}

\bibitem[{{De Sanctis} {et~al.}(2015){De Sanctis}, {Capaccioni}, {Ciarniello},
  {Filacchione}, {Formisano}, {Mottola}, {Raponi}, {Tosi},
  {Bockel{\'e}e-Morvan}, {Erard}, {Leyrat}, {Schmitt}, {Ammannito}, {Arnold},
  {Barucci}, {Combi}, {Capria}, {Cerroni}, {Ip}, {Kuehrt}, {McCord}, {Palomba},
  {Beck}, {Quirico}, {VIRTIS Team}, {Piccioni}, {Bellucci}, {Fulchignoni},
  {Jaumann}, {Stephan}, {Longobardo}, {Mennella}, {Migliorini}, {Benkhoff},
  {Bibring}, {Blanco}, {Blecka}, {Carlson}, {Carsenty}, {Colangeli}, {Combes},
  {Crovisier}, {Drossart}, {Encrenaz}, {Federico}, {Fink}, {Fonti}, {Irwin},
  {Langevin}, {Magni}, {Moroz}, {Orofino}, {Schade}, {Taylor}, {Tiphene},
  {Tozzi}, {Biver}, {Bonal}, {Combe}, {Despan}, {Flamini}, {Fornasier},
  {Frigeri}, {Grassi}, {Gudipati}, {Mancarella}, {Markus}, {Merlin}, {Orosei},
  {Rinaldi}, {Cartacci}, {Cicchetti}, {Giuppi}, {Hello}, {Henry}, {Jacquinod},
  {Rees}, {Noschese}, {Politi}, \& {Peter}}]{desanctis2015}
{De Sanctis}, M.~C., {Capaccioni}, F., {Ciarniello}, M., {et~al.} 2015, Nature,
  525, 500

\bibitem[{{Fray} \& {Schmitt}(2009)}]{fray2009}
{Fray}, N. \& {Schmitt}, B. 2009, Planetary and Space Science, 57, 2053

\bibitem[{{Goesmann} {et~al.}(2015){Goesmann}, {Rosenbauer}, {Bredeh\"oft},
  {Cabane}, {Ehrenfreund}, {Gautier}, {Giri}, {Kr\"uger}, {Le Roy},
  {MacDermott}, {McKenna-Lawlor}, {Meierhenrich}, {Mu\~noz Caro}, {Raulin},
  {Roll}, {Steele}, {Steininger}, {Sternberg}, {Szopa}, {Thiemann}, \&
  {Ulamec}}]{goesmann2015}
{Goesmann}, F., {Rosenbauer}, H., {Bredeh\"oft}, J.-H., {et~al.} 2015, Science,
  349, aab0689

\bibitem[{{Goesmann} {et~al.}(2007){Goesmann}, {Rosenbauer}, {Roll}, {Szopa},
  {Raulin}, {Sternberg}, {Israel}, {Meierhenrich}, {Thiemann}, \&
  {Munoz-Caro}}]{goesmann2007}
{Goesmann}, F., {Rosenbauer}, H., {Roll}, R., {et~al.} 2007, Space Science
  Reviews, 128, 257

\bibitem[{{H{\"a}ssig} {et~al.}(2015){H{\"a}ssig}, {Altwegg}, {Balsiger},
  {Bar-Nun}, {Berthelier}, {Bieler}, {Bochsler}, {Briois}, {Calmonte}, {Combi},
  {De Keyser}, {Eberhardt}, {Fiethe}, {Fuselier}, {Galand}, {Gasc}, {Gombosi},
  {Hansen}, {J{\"a}ckel}, {Keller}, {Kopp}, {Korth}, {K{\"u}hrt}, {Le Roy},
  {Mall}, {Marty}, {Mousis}, {Neefs}, {Owen}, {R{\`e}me}, {Rubin}, {S{\'e}mon},
  {Tornow}, {Tzou}, {Waite}, \& {Wurz}}]{haessig2015}
{H{\"a}ssig}, M., {Altwegg}, K., {Balsiger}, H., {et~al.} 2015, Science, 347,
  aaa0276

\bibitem[{{H\"ubner} {et~al.}(2006){H\"ubner}, {Benkhoff}, {Capria},
  {Coradini}, {De Sanctis}, {Orosei}, \& {Prialnik}}]{huebner2006}
{H\"ubner}, W.~F., {Benkhoff}, J., {Capria}, M.-T., {et~al.} 2006, {Heat and
  Gas Diffusion in Comet Nuclei} (The International Space Science Institute,
  Bern, Switzerland, by ESA Publications Division, Noordwijk, The Netherlands,
  SR-004)

\bibitem[{{Kr{\"u}ger} {et~al.}(2015){Kr{\"u}ger}, {Seidensticker}, {Fischer},
  {Albin}, {Apathy}, {Arnold}, {Flandes}, {Hirn}, {Kobayashi}, {Loose},
  {P{\'e}ter}, \& M.}]{krueger2015b}
{Kr{\"u}ger}, H., {Seidensticker}, K.~J., {Fischer}, H.~H., {et~al.} 2015,
  Astronomy and Astrophysics, 583, A15

\bibitem[{{Mart{\'{\i}}n-Dom{\'e}nech}
  {et~al.}(2014){Mart{\'{\i}}n-Dom{\'e}nech}, {Mu{\~n}oz Caro}, {Bueno}, \&
  {Goesmann}}]{martin-domenech2014}
{Mart{\'{\i}}n-Dom{\'e}nech}, R., {Mu{\~n}oz Caro}, G.~M., {Bueno}, J., \&
  {Goesmann}, F. 2014, Astronomy and Astrophysics, 564, A8

\bibitem[{{Morse} {et~al.}(2015){Morse}, {Wright}, \& {et al.}}]{morse2015}
{Morse}, A., {Wright}, I., \& {et al.}, A. 2015, Astronomy and Astrophysics,
  583, A42

\bibitem[{{Prialnik} {et~al.}(2005){Prialnik}, {Benkhoff}, \&
  {Podolak}}]{prialnik2005}
{Prialnik}, D., {Benkhoff}, J., \& {Podolak}, M. 2005, in Comets II, ed. M.~C.
  {Festou}, H.~U. {Keller}, \& H.~A. {Weaver} (University of Arizona Press),
  359--387

\bibitem[{{Roll} \& {Witte}(2016)}]{roll2016b}
{Roll}, R. \& {Witte}, L. 2016, Planetary and Space Science, 125, 12

\bibitem[{{Roll} {et~al.}(2016){Roll}, {Witte}, \& {Arnold}}]{roll2016a}
{Roll}, R., {Witte}, L., \& {Arnold}, W. 2016, Icarus, 280, 359

\bibitem[{{Spohn} {et~al.}(2015){Spohn}, {Knollenberg}, {Ball}, {Banaskiewicz},
  {Benkhoff}, {Grott}, {Grygorczuk}, {H\"uttig}, {Hagermann}, {Kargl},
  {Kaufmann}, {K\"omle}, {K\"uhrt}, {Kossacki}, {Marczewski}, {Pelivan},
  {Schr\"odter}, \& {Seiferlin}}]{spohn2015}
{Spohn}, T., {Knollenberg}, J., {Ball}, A.~J., {et~al.} 2015, Science, 349,
  aab0464

\bibitem[{{Stein}(2015)}]{stein2015}
{Stein}, S.~E. 2015, in {NIST Chemistry WebBook}, ed. P.~J. {Linstrom} \& W.~G.
  {Mallard}, http://webbook.nist.gov

\bibitem[{{Wright} {et~al.}(2007){Wright}, {Barber}, {Morgan}, {Morse},
  {Sheridan}, {Andrews}, {Maynard}, {Yau}, {Evans}, {Leese}, {Zarnecki},
  {Kent}, {Waltham}, {Whalley}, {Heys}, {Drummond}, {Edeson}, {Sawyer},
  {Turner}, \& {Pillinger}}]{wright2007}
{Wright}, I.~P., {Barber}, S.~J., {Morgan}, G.~H., {et~al.} 2007, Space Science
  Reviews, 128, 363

\bibitem[{{Wright} {et~al.}(2015){Wright}, {Sheridan}, {Barber}, {Morgan},
  {Andrews}, \& {Morse}}]{wright2015}
{Wright}, I.~P., {Sheridan}, S., {Barber}, S.~J., {et~al.} 2015, Science, 349,
  aab0673

\bibitem[{{Yamamoto} {et~al.}(1983){Yamamoto}, {Nakagawa}, \&
  {Fukui}}]{yamamoto1983}
{Yamamoto}, T., {Nakagawa}, N., \& {Fukui}, Y. 1983, Astronomy and
  Astrophysics, 122, 171

\end{thebibliography}

\end{document}